\documentclass{pasj00}
\usepackage[dvips]{graphicx}
\usepackage{natbib}
\draft

\begin{document}
\SetRunningHead{N.Nakasato}{}
\Received{}
\Accepted{}

\title{SPH Simulations with Reconfigurable Hardware Accelerator}

\author{Naohito \textsc{Nakasato}, \altaffilmark{1}
Tsuyoshi \textsc{Hamada}, \altaffilmark{1,2}
and
Toshiyuki \textsc{Fukushige} \altaffilmark{2,3}
}

\altaffiltext{1}{Computational Astrophysics Laboratory,\\ 
The Institute of Physical and Chemical Research,\\ 
2-1 Hirosawa, Wako, Saitama, 251-0198}

\altaffiltext{2}{Department of General System Studies,\\
College of Arts and Sciences, The University of Tokyo,\\
3-8-1 Komaba, Meguro, Tokyo 153-8902} 

\altaffiltext{3}{Tokyo Laboratory,\\ 
K\&F Computing Research Co.,\\
840-205 Arai, Hino, Tokyo 191-0022}

\email{nakasato@riken.jp, thamada@riken.jp, fukushig@provence.c.u-tokyo.ac.jp}


\maketitle

\begin{abstract}
We present a novel approach to accelerate astrophysical hydrodynamical simulations.
In astrophysical many-body simulations, GRAPE (GRAvity piPE) system has been widely used
by many researchers. 
However, in the GRAPE systems, its function is completely fixed because 
specially developed LSI is used as a computing engine.
Instead of using such LSI, we are developing a special purpose computing system
using Field Programmable Gate Array (FPGA) chips as the computing engine.
Together with our developed programming system, we have implemented computing pipelines
for the Smoothed Particle Hydrodynamics (SPH) method on our PROGRAPE-3 system.
The SPH pipelines running on PROGRAPE-3 system have the peak speed of 85 GFLOPS 
and in a realistic setup, the SPH calculation using one PROGRAPE-3 board is
5-10 times faster than the calculation on the host computer.
Our results clearly shows for the first time
that we can accelerate the speed of the SPH simulations of a simple astrophysical phenomena
using considerable computing power offered by the hardware.
\end{abstract}

\section{Introduction}
GRAPE \citep[``GRAvity piPE'';][]{Sugimoto_1990, Makino_1998} is a special-purpose
computer originally invented and developed for simulations of 
evolution of a star cluster.
In such simulations, most time consuming part of the simulation
is calculation of gravity force between stars (or pseudo-particles).
After the first development of GRAPE-1, many generation of GRAPE computers
have been developed and used very widely for
simulations of star cluster \citep[e.g.,][]{Makino_1996, Portegies_2004}, 
galaxy evolution \citep[e.g.,][]{Steinmetz_1996, Mori_1999, Nakasato_2003},
cluster of galaxies \citep[e.g.,][]{Sensui_1999, Fukushige_2001, Fukushige_2004},
evolution of planetesimal \citep[e.g.,][]{Kokubo_1996}, 
formation of Moon \citep[e.g.,][]{Kokubo_2000}, 
and many more. For details see reviews \citep{Makino_1998, Hut_1999, Makino_2003a}. 

In most of GRAPEs, calculation of gravitational interaction between particles expressed as
\begin{equation}
\mbox{\boldmath{$f$}}_{i} =
	 -\sum_{j=1}^{N} m_j\frac{\mbox{\boldmath{$r$}}_i - \mbox{\boldmath{$r$}}_j}
		{(|\mbox{\boldmath{$r$}}_i - \mbox{\boldmath{$r$}}_j|^2 + \epsilon^2)^{3/2}},
\label{g}
\end{equation}
is implemented as fully pipelined logic on a LSI (a.k.a. GRAPE chip).
Here $\mbox{\boldmath{$r$}}_i$ and $m_i$ are the position and mass of $i$-th particle,
and $\epsilon$ is the softening parameter.
Its relatively less complex formulation
is one of reasons that GRAPE hardwares can offer huge advantage
both in the absolute speed and in the price-performance. 

Although many astrophysical phenomena are
primarily driven by gravitational force,
hydrodynamics and other physics are sometimes important.
A number of research groups use GRAPE hardwares mainly for combined
$N$-body+hydrodynamical simulations such as
the evolution of galaxies or clusters galaxies.
\citep[e.g.,][]{Steinmetz_1996, Mori_1999, Nakasato_2003}.
Those groups including ourselves use Smoothed Particles Hydrodynamics (SPH) method
\citep{Lucy_1977, Gingold_1977} for hydrodynamical part of the codes.
In the SPH method, fluid is expressed as bunch of particles and
hydrodynamical interaction such as pressure force is calculated with a following
summation between particles;
\begin{equation}
\mbox{\boldmath{$f$}}_{i} =
- \sum m_j \left( \frac{P_i}{\rho^2_i} + \frac{P_j}{\rho^2_j} + \Pi_{ij} \right)
\nabla W(\mbox{\boldmath{$r$}}_i - \mbox{\boldmath{$r$}}_j;h_{ij}), 
\label{a}
\end{equation}
where $\rho_i$, $P_i$ are density and pressure for $i$-th particle, respectively, 
and $h_{ij}$ is the average smoothing length between 
$i$-th particle and $j$-th particle.
and $W(r;h)$ is the kernel function for weighting physical quantities.
Furthermore, $\Pi_{ij}$ is an artificial viscosity term to preserve
better numerical stability.
In its abstract from, gravity interaction and SPH interaction have a similar structure.
However, hydrodynamical interaction is short-range, 
and summation for $i$-th particle in equation (\ref{a}) is done over neighbor particles
whose distance from $i$-th particle is less than $2h$.
Accordingly, calculation cost of SPH interaction is $O(n N)$
instead of $O(N^2)$,
where $n$ is an average number of neighbor particles,
typically 30-100 in 3 dimensional simulations.

The previous implementations of SPH on GRAPE use it only to calculate gravitational
interaction and to construct a list of neighbors for the SPH interactions.
Actual evaluation of the SPH interaction (e.g., equation (\ref{a})) is performed
on the host computer. Thus, in the SPH simulations with GRAPE,
the speed of the host computer tends to determine the total performance.
This is because the calculation cost of the SPH interaction is
much larger than the calculation cost of the rest of the simulation program
such as time integration and I/O
though not as large as that of the gravitational interaction.
The calculation cost of single SPH interaction
is a few times more than that of the gravitational interaction, 
e.g., calculation of one SPH interaction needs about 160 floating operations whereas
one gravity interaction needs 38 floating operations.
If we perform pure SPH simulations without collisionless particles
(i.e., interacting with other particles only through gravity),
the gain in speed achieved by GRAPE is a factor of few at the best.
Practically, in simulations of a galaxy,
relatively large number of collisionless particles
must be included because of large mass ratio between dark matter
and baryon in the universe, and for such calculations GRAPE can offer a large speedup.
However, when we want to perform the SPH simulations without collisionless 
particles, the speedup would be rather limited. Accordingly, though the SPH
techniques have been applied to wide variety of problems, such as the
dynamics of star-forming regions and hydrodynamical interaction of
stars, GRAPE hardwares have not been widely used for those kind of
simulations.

In principle, one could develop a special-purpose hardware similar to
GRAPE to accelerate the SPH interactions \citep{Yokono_1999}.
Such a hardware, once completed, would offer a large speedup over general-purpose computers. 
However, so far feasibility of such a project has still remained unproven.
There are several reasons why it is
difficult to develop a specialized computer for the SPH method.
First, the calculation of the SPH interaction is a bit more complex
compared to the calculation of the gravitational interaction as already shown.
In concept, implementation of {\it SPH chip} is equally simple to GRAPE chip.
What we need is a hardware to evaluate sum of quantities weighted
by the kernel function $W$ or its derivative.
In practice, there are numerous technical details which have to be taken care of.
For example, the smoothing length $h$ is different for different particles, and
$W$ has to be symmetrized to preserve momentum conserving nature of the scheme.

Another difficulty is that in SPH algorithms \citep{Monaghan_1992}
there are rather large number of varieties.
A crucial variety is how to symmetrize $W$.
In addition, there are many methods to implement artificial viscosity $\Pi$.
Moreover, there have been several new developments in the last
few years \cite[e.g.,][]{Springel_2002}.
In other words, a number of {\it numerical parameters} that control and change
details of SPH interaction is much larger (say 10 or more)
than the gravity interaction which has practically one parameter $\epsilon$.
If we would design the SPH chip, 
the development of the hardware would be significantly more difficult and
time-consuming compared to the development of the GRAPE chip,
and yet the hardware might become obsolete even before its completion.

We can avoid the risk of becoming obsolete if we can change the
hardware of SPH chip {\it after its completion}.
Changing the hardware might sound self-contradictory,
since the hardware is, unlike the software, cannot be changed once it is completed.
Though the name might sound strange, such
``programmable'' chips have been available for several decades now.
These chips, usually called FPGA (Field-Programmable Gate Array) chips,
consist of small logic elements (LE) and switching matrix to connect them.
One LE is typically a small lookup table made of SRAM,
combined with additional circuits such as a flip-flop and special logic for arithmetic operations. 

The size (in equivalent gate count) of the FPGA chips has been enormously
increased, from around 1k of mid-1980's to around 10 million of 2005.
Since the increase is driven essentially by the advance in the
semiconductor device technology, one can expect that the increase will
continue for the next decade. 
An FPGA chip with over 20 million gates will be available by 2006.
Those FPGA chips are large enough to house fairly complex circuits.
For example, the pipeline processor chip of GRAPE-3 \citep{Okumura_1993} is around 20k gates.
Thus, even if we assume rather low utilization ratio of 20\%, 
one GRAPE-3 chip should fit into an FPGA chip with 100k gates, 
which is now quite typical and pretty cheap.

In 1990's, to implement GRAPE pipelines on FPGA would have little practical
meaning, since one GRAPE chip of similar price to one FPGA chip 
at that time was much faster.
However, current situation is the other way around
such that for relatively small budget project, 
cost to design and implement new GRAPE chip even with current
modest technology is rather expensive
(i.e., the initial development cost of GRAPE-6 chip has exceeded
1M USD; \cite{Makino_2003b}) and it is not necessarily
true that with fixed amount of money, 
GRAPE chips are cheaper and faster than FPGA chips with GRAPE pipelines.
Thus, to implement GRAPE pipelines on FPGA itself is now becoming
a considerable alternative to make a new GRAPE chip.
Moreover, applications for which no custom hardware is available
would find a GRAPE-like hardware implemented using FPGA useful.
The SPH interaction is a perfect example for this type of hardware.
We may implement any variety of the SPH algorithm, as far as it is expressed
as interaction between particles and the necessary circuit fits in a
target FPGA chip.
The peak performance is not as large as what we can
enjoy with custom LSI chips, but could be still orders of magnitude faster
than what is available on general-purpose computers.

We call the concept of using a programmable FPGA chip as the pipeline
processor for astrophysical simulations 
as PROGRAPE (PROgrammable GRAPE) that has been introduced by \cite{Hamada_2000}.
It should be noted that PROGRAPE is not the first attempt to use FPGAs as the building blocks
of special-purpose computers.  There are a number of projects to
develop custom computing machines using FPGAs as a main unit.
The most influential one are Splash-1 and Splash-2 project \citep{Buell_1996}.

In the present work, we report current status of our project that
constructs a new generation of FPGA-based GRAPE-like hardware
and tries to implement SPH pipeline on the hardware.
We have developed our third-generation of PROGRAPE type hardware which we call PROGRAPE-3.
We mainly concentrate to report 
about details of implementation of SPH pipelines on PROGRAPE-3
and its performance obtained so far.
In the companion paper \citep{Hamada_2006},
we will report details of implementation
and its performance of GRAPE pipelines on PROGRAPE-3 and other FPGA-base hardware.
Ultimately, in the present work, 
we prove our concept of PROGRAPE is now feasible and very promising.
In fact our result clearly shows that 
at least for astrophysical simulations, 
{\it the reconfigurable supercomputing era} is coming now.

This paper is organized as follows. In section 2, we briefly introduce
our third generation of a reconfigurable hardware accelerator
for astrophysical simulations. For our new computing board,
we have developed special software to implement processors on the board.
Using the developed software, we presents details of what we implement
on our board in section 3. Here, we describe how a SPH simulation code
can be fitted to our reconfigurable hardware. 
And more importantly, we present how large is the performance improvement obtained so far
for SPH simulations. 
And finally, we summarize our current status and present some possible
improvements of our software/hardware system.

\section{PROGRAPE-3 system: new reconfigurable hardware accelerator}
\begin{figure}
\begin{center}
\scalebox{0.5}{\includegraphics{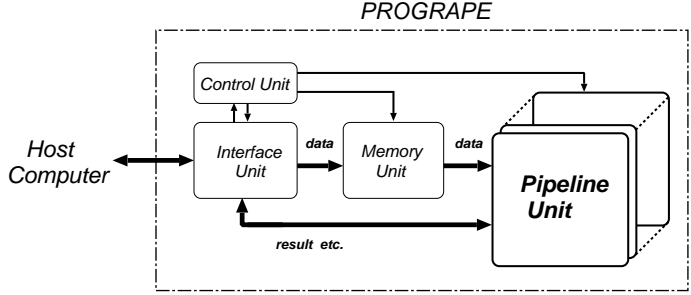}}
\caption{Schematic structure of PROGRAPE system.}
\label{PROGRAPE}
\end{center}
\end{figure}

In this section, we briefly describe the hardware part of our system.
As a sub-project of the GRAPE project, PROGRAPE-1 \citep{Hamada_2000},
and PROGRAPE-2 have been developed so far.
Conceptually, both systems have the same structure as schematically shown in 
Figure \ref{PROGRAPE}.
PROGRAPE system consists of a host computer and a PROGRAPE board.
This structure is exactly the same as conventional GRAPE systems
and its operational scheme is also same,
i.e., the PROGRAPE board calculates interactions between particles and the host computer 
performs all other complex calculations.
Inside the PROGRAPE board, there are four units
(interface unit, control unit, memory unit and computational pipeline unit) 
as shown in Figure \ref{PROGRAPE}, and this structure is basically also the same
as conventional GRAPE boards.
The Interface unit is used for communication between the host computer and the board through PCI bus.
The Pipeline unit is central engine of the system that calculates interactions between particles.
A crucial difference between GRAPE and PROGRAPE is that in the conventional GRAPE,
computational pipeline is implemented on specially developed LSIs and hence its function is fixed,
on the other hand, the pipeline is implemented on FPGA chips in the case of PROGRAPE.
Furthermore, those pipelines on the FPGA chips can be reconfigurable or programmable.
This programmability is a most drastic difference between conventional GRAPE and PROGRAPE.

\subsection{PROGRAPE-3 system details}
\begin{figure}
\begin{center}
\scalebox{1.0}{\includegraphics{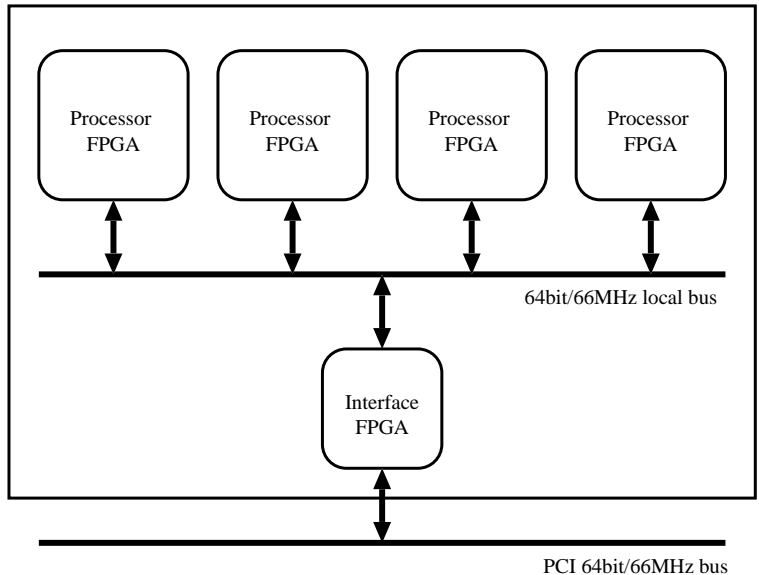}}
\caption{Schematic structure of the Bioler-3 board.}
\label{Bioler3}
\end{center}
\end{figure}

PROGRAPE-3 system is the third generation of the PROGRAPE architecture.
The system consists of a host computer and single(multiple) FPGA board(s).
For our target FPGA board in the present work, we adopt Bioler-3 board that
has been developed as a joint project between 
Chiba University and RIKEN.
The Bioler-3 board is a programmable multi-purpose computer board 
dedicated to accelerate computationally intensive applications.
Here, we use the Bioler-3 board for implementing the computational pipeline for
astrophysical many-body simulations as described later.
In Figure \ref{Bioler3}, we show main components implemented on the Bioler-3 board.
It has four large FPGA chips (hereafter, processor FPGA chip) dedicated for the central engine 
and one small FPGA chip (interface FPGA chip) for interface unit.
Also, the control unit and the memory unit are implemented on the processor FPGA chips.
Note in PROGRAPE-1 board and all version of the GRAPE board,
memory chips are mounted on those boards for the memory unit.
We use a PC whose CPU is AMD Opteron 2.2/2.4 GHz as the host computer of the PROGRAPE-3 system.
The Bioler-3 board and the host computer is connected through PCI 64-bit/66MHz bus 
that has theoretical bandwidth of 512 MB s$^{-1}$.

In a rather simplified expression, all PROGRAPE system so far is optimized 
to evaluate a following formula:
\begin{equation}
\mbox{\boldmath{$f$}}_{i} = \sum_{j=1}^{N} \mbox{\boldmath{$G$}} (\mbox{\boldmath{$X$}}_i, \mbox{\boldmath{$X$}}_j), 
\label{basic}
\end{equation}
where $\mbox{\boldmath{$X$}}_i$ is a vector contains information related to the $i$-th particle,
$\mbox{\boldmath{$f$}}_i$ is the resulted {\it force} on the $i$-th particle
and $\mbox{\boldmath{$G$}}$ is a function which describes the interaction
between the $i$-th and the $j$-th particle.
To give a specific example, in the case of gravitational interaction, 
$\mbox{\boldmath{$X$}}_i = (\mbox{\boldmath{$x$}}_i, m_i)$ and $\mbox{\boldmath{$G$}}$
is given as equation (\ref{g}).
Note that $\mbox{\boldmath{$X$}}_i$ and $\mbox{\boldmath{$X$}}_j$ in equation (\ref{basic}) do
not necessarily contain same amount of information, though we refer to them by the same symbol for simplicity.
In the case of gravitational interaction, mass $m_i$ is not used to evaluate equation (\ref{basic}).
Therefore, we can remove $m_i$ from $\mbox{\boldmath{$X$}}_i$ through $\mbox{\boldmath{$X$}}_j$ 
should contain $m_j$.
If PROGRAPE is used to evaluate SPH interaction,
$\mbox{\boldmath{$X$}}_i$ should contain density and pressure to calculate equation (\ref{a}),
and velocity and other physical quantities to calculate $\Pi$. 
And in this case, function $\mbox{\boldmath{$G$}}$ would be given in equation (\ref{a}). 
Details will be described in a later section.

The calculation with PROGRAPE proceeds in the following steps.

(1) The host computer sends a configuration data, 
that is a binary code for FPGA chips, to the processor FPGAs
for configuring the FPGA chips to do what one wants to calculate.
The configuration data contain all necessary information 
such that how the control and memory unit is organized and
what the pipeline unit calculates etc. This step needs to be done only once.

(2) The host computer sends the data of particles to the memory unit.
These data correspond to $\mbox{\boldmath{$X$}}_j$ in equation (\ref{basic}), 
and we refer to them as ``$j$-data''.

(3) After sending $j$-data, the host computer sets $\mbox{\boldmath{$X$}}_i$
to the pipeline unit. These data ($i$-data) are stored in registers in the pipeline unit, 
which we call ``$i$-register''. Note if the pipeline unit has multiple pipelines
inside, each pipeline has own ``$i$-register'' and the host computer should set
multiple $i$-data to each pipeline. 

(4) Then, the host computer sends the command to the control unit to start the calculation.
Accordingly, the control unit starts to send indexes (address) to the memory,
and sends start signal to the pipeline unit. At the same time, the memory unit
receives the index at each clock and feeds corresponding $j$-data to the pipeline unit.
The pipeline unit receives $j$-data and evaluates
interaction between those $j$-data and own $i$-data and accumulate results
in its internal accumulators.

(5) Finally, when the summation is finished, the host computer receives a signal
and reads the accumulated results ($f$-data) from the pipeline unit.

Note that in the PROGRAPE architecture, one can change only the configuration data of
the processor FPGA. That is one can implement an arbitrary interaction function
$\mbox{\boldmath{$G$}}$ 
in the pipeline unit and change the memory and control units if necessary.
For example, one would change the amount of $j$-data transfered from the memory 
unit to the pipeline unit in one clock to optimize data transfer performance.
We are developing a software package to implement or create 
the content of the processor FPGA chip as described in the next section.

\begin{figure}
\begin{center}
\scalebox{0.3}{\includegraphics{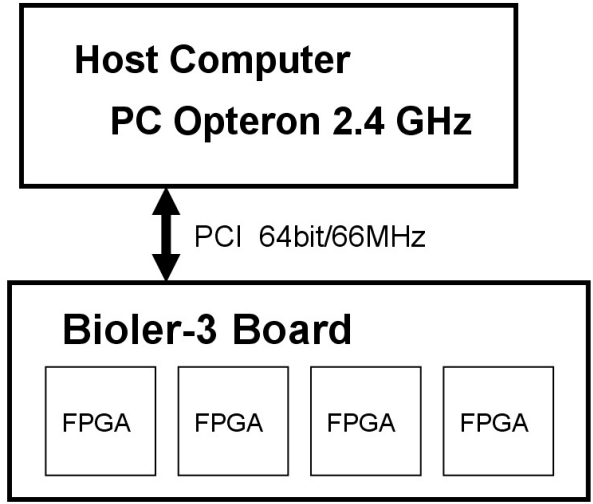}}
\caption{
The present configuration of PROGRAPE-3 system.
}
\label{PROGRAPE3}
\end{center}
\end{figure}

The processor FPGA chips used in the present work are Xilinx XC2VP70-5 chip.
This FPGA chip has 66,176 logic cells, 328 16Kbit-SRAM blocks, 328 multiplier blocks
with the size of 18-bit. The manufacturer claims that the chip has the capacity equivalent
to about 10 million gates.
As already noted the pipeline processor chip of GRAPE-3 needs only 20k gates.
Even with one processor FPGA chip, one can expect to implement plenty of complex 
particle interaction function on it.
In Figure \ref{PROGRAPE3}, we present schematic structure of PROGRAPE-3 system. 

\subsection{Software architecture for PROGRAPE-3 system}
The way one use PROGRAPE system is almost exactly the same
as the way we use any of GRAPE systems
if one already had the configuration data for the processor FPGA.
A distinction between PROGRAPE and GRAPE is that in the case of PROGRAPE, 
the user {\it can} specify every details of the pipeline unit with the configuration data.
To change its logic in a slightly different way, the user {\it must} 
specify literally every bit of the configuration data for the processor FPGA.
To generate such configuration data, one must know how to program FPGA and
a difficult thing is that the FPGA programming is not similar to 
or completely different from conventional programming in Fortran or C language.
There are several reasons for this difficulty.
Apparent obstacle is that there are too much freedom.
An FPGA chips can be used to implement any logic circuit , as far as it fits into the chip.
Thus, the FPGA programming is essentially to specify every detail of the logic
using very basic logic gates (such as AND/OR gates) and flip-flops.
To make an analogy, an FPGA is something like a universal Turing machine
without any high-level languages, library functions or even operating systems.

At present, those who wish to use PROGRAPE for their own work must 
create configuration data for the processor FPGA 
by writing its design with any of hardware description languages (such as VHDL or Verilog), 
unless the application has been developed by somebody else.
Although there are a several commercial software available that make the FPGA
programming/designing easy, such software usually has only limited applicability.
We by ourself are now developing a methodology that would greatly reduce amount
of development time and cost for users of PROGRAPE.
Our software PGR (Processors Generators for Reconfigurable systems) \citep{Hamada_2005a}
is a newest version of former programing system PGPG software \citep{Hamada_2005b}.
Although PGR has a several improvements over PGPG software, 
a most crucial one is that PGR supports floating-point arithmetic operations
with variable bit length.

Support of the floating-point arithmetic is crucial to implement SPH interactions
on PROGRAPE system as explained below.
So far the first GRAPE system (GRAPE-1)
and its replacement GRAPE systems (GRAPE-3 and 5) 
have used the logarithmic number system (LNS) to implement the gravity pipeline.
In LNS, a real number is converted to logarithm of the number.
Accordingly, multiply, division, and square operations in LNS 
are done with addition, subtraction and shift operations, respectively, 
and in compensation addition and subtraction operations
must be done with a table-lookup function 
\citep[for details, see][]{Kawai_2000}.
In the case of the GRAPE pipeline, 
we need 8 addition/subtraction,
5 multiply, 1 division and 3 square operations for computing equation (\ref{g}).
Since implementation of multiply and division operation as logic circuit needs
much more resources than addition or subtraction,
this nature of LNS is highly desirable in terms of amount of needed resource.
Therefore, in the PGPG project, 
they have only supported LNS and integer number operations.
For traditional FPGAs, 
it is true that implementation of multiply needs larger amount of resources.
However, recent FPGA chips including the chips we used in the present work
have dedicated special logic for multiply operations.
Those special logic are desirable for applications that need many multiply operations.
On the other hand, we also need a plenty of addition and subtraction
to implement SPH pipelines as shown in later sections.
In such situation, LNS has no big advantage over
conventional floating point number representation.
Furthermore, floating point number representation is more accordant
than LNS to those who are programming in Fortran or C language.

Essential ingredients of our methodology is
that from a description of a pipeline written in a new language,
our software PGR automatically generates or creates (1) necessary VHDL source codes for
the configuration file of the processor FPGA,
(2) application program interface for a user application, and
(3) a software bit-level emulator that is necessary to test whether one description
of the pipeline is correct or not.
Details on PGR are described in \cite{Hamada_2005a} and 
more specific details on implementation and performance results
of GRAPE pipeline developed using PGR have been
discussed in the companion paper \citep{Hamada_2006}.
In the course of development for the present work, 
we have extensively used and tested PGR to implement 
SPH pipelines on PROGRAPE-3 system.

\section{SPH implementation on Reconfigurable Hardware Accelerator}
In the SPH method, pressure force or any other physical variables are expressed as 
particle interaction and its summation over neighbor particles.
In the SPH simulations, we integrate the equation of motion for particles 
to evolve a system that we want to simulate or model.
There are huge number of literatures for various formulation of the SPH method
(see a review for mainly astrophysical applications \cite{Monaghan_1992} 
and for non-astrophysical applications \cite{Liu_2003}).
In the present work, since our main interest is the evolution of galaxies
and we have our own SPH code \citep{Nakasato_2003} for mainly modeling the evolution of galaxies, 
we have adopted widely used formulation for galaxy evolution models
as follows \citep{Navarro_1993}.
Here, we present equations what we exactly implement as pipelines for PROGRAPE-3 system.
In the followings, the kernel function we adopted is usual spline kernel proposed
by \cite{Monaghan_1985} as follows:
\begin{equation}
 W(q, h) = \frac{1}{\pi h^3}
  \left\{
   \begin{array}{cl}
    1 - \frac{3}{2}q^2 + \frac{3}{4}q^3, & \quad (0 \le q \le 1)\\
    \frac{1}{4}(2 - q)^2,                & \quad (1 \le q \le 2),\\
    0,                                   & otherwise
   \end{array}
  \right\}
\label{kernel}
\end{equation}
where $q$ is defined as $q = r/h$. 

\subsection*{density: $\rho$}
\begin{equation}
\rho_i = \sum m_j W(\mbox{\boldmath{$r$}}_i-\mbox{\boldmath{$r$}}_j;h_{ij}).
\label{rho}
\end{equation}

\subsection*{divergence of velocity: $\nabla \cdot \mbox{\boldmath{$v$}}$}
\begin{equation}
\rho_i (\nabla \cdot \mbox{\boldmath{$v$}})_i = 
\sum m_j (\mbox{\boldmath{$v$}}_j-\mbox{\boldmath{$v$}}_i) \cdot \nabla W(\mbox{\boldmath{$r$}}_i-\mbox{\boldmath{$r$}}_j;h_{ij}).
\label{v1}
\end{equation}

\subsection*{rotation of velocity: $\nabla \times \mbox{\boldmath{$v$}}$}
\begin{equation}
\rho_i (\nabla \times \mbox{\boldmath{$v$}})_i =
\sum m_j (\mbox{\boldmath{$v$}}_j-\mbox{\boldmath{$v$}}_i) \times \nabla W(\mbox{\boldmath{$r$}}_i-\mbox{\boldmath{$r$}}_j;h_{ij}).
\label{v2}
\end{equation}

\subsection*{estimate of a number of neighbors: $n_{\rm neighbor}$}
\begin{equation}
{\rm n}_i = \sum W_{\rm neighbor}(\mbox{\boldmath{$r$}}_i-\mbox{\boldmath{$r$}}_j;h_{ij}).
\label{nn}
\end{equation}

\subsection*{pressure force: $\dot {\mbox{\boldmath{$v$}}}$}
\begin{equation}
\dot {\mbox{\boldmath{$v$}}}_i =  
- \sum m_j \left( \frac{P_i}{\rho^2_i} + \frac{P_j}{\rho^2_j} + \Pi_{ij} \right)
\nabla W( \mbox{\boldmath{$r$}}_i - \mbox{\boldmath{$r$}}_j;h_{ij}).
\label{pf}
\end{equation}

\subsection*{time derivative of energy: $\dot u$}
\begin{equation}
\dot u_i = 
\sum m_j \left( \frac{P_i}{\rho^2_i} + \frac{1}{2} \Pi_{ij} \right) (\mbox{\boldmath{$v$}}_i - \mbox{\boldmath{$v$}}_j) \cdot \nabla W(\mbox{\boldmath{$r$}}_i - \mbox{\boldmath{$r$}}_j;h_{ij}).
\label{u}
\end{equation}

\subsection*{artificial viscous term: $\Pi$}
\begin{equation}
\Pi_{ij} =
  \left\{
   \begin{array}{cl}
    f_{ij} \frac{-\alpha c_{ij} \mu_{ij} + \beta \mu^2_{ij}}{\rho_{ij}},
     & \quad \mbox{if} \quad
     (\mbox{\boldmath{$v$}}_i - \mbox{\boldmath{$v$}}_j) \cdot
     (\mbox{\boldmath{$r$}}_i - \mbox{\boldmath{$r$}}_j) \le 0 \\
    0, & \quad \mbox{otherwise},\\
   \end{array}
  \right\},
\label{vis}
\end{equation}
\begin{equation}
f_{ij} = \frac{f_i + f_j}{2}, 
\end{equation}
\begin{equation}
f_{i} = \frac{|\nabla \cdot \mbox{\boldmath{$v$}}|}{|\nabla \cdot \mbox{\boldmath{$v$}}| + \nabla \times \mbox{\boldmath{$v$}} + 10^{-4} c_i/h_i}
\label{ff}
\end{equation}
\begin{equation}
\mu_{ij} =
 \frac{h_{ij} (\mbox{\boldmath{$v$}}_i - \mbox{\boldmath{$v$}}_j)\cdot
 (\mbox{\boldmath{$r$}}_i - \mbox{\boldmath{$r$}}_j)}
 {(\mbox{\boldmath{$r$}}_i - \mbox{\boldmath{$r$}}_j)^2 + (0.1 h_{ij})^2}.
\label{uu}
\end{equation}
Here, $c$ is sound velocity of the particle and
$h_{ij}$, $c_{ij}$ and $\rho_{ij}$ represent
average values between $i$-th particle and $j$-th particle, 
namely, $h_{ij} = (h_i + h_j)/2$ etc.
Since equation (\ref{nn}) is not commonly used in the conventional SPH scheme,
we note that this estimate of a number of neighbors is used to update 
smoothing length $h$ according to an algorithm proposed by \cite{Thacker_2000}
and a modified kernel function $W_{\rm neighbor}$ (see equation (9) in \cite{Thacker_2000}
for definition) is used for this purpose.

Currently, we implement those equations as two separate stages as described below.
\begin{itemize}
\item first stage:\\
We compute equations (\ref{rho}), (\ref{v1}), (\ref{v2}), and (\ref{nn}).
This is because we need $\rho$, $\nabla \cdot \mbox{\boldmath{$v$}}$ and $\nabla \times \mbox{\boldmath{$v$}}$
for computing equations (\ref{pf}) and (\ref{u}) through artificial viscous term in equations in
(\ref{vis})-(\ref{uu}).
\item second stage:\\
We compute equations (\ref{pf}) and (\ref{u}).
During computation of those values, we also calculate equations in (\ref{vis})-(\ref{uu}).
\end{itemize}
Accordingly, for the first stage, 
data vector $\mbox{\boldmath{$X$}}_i$ ($i$-data) contains
position $\mbox{\boldmath{$r$}}$, velocity $\mbox{\boldmath{$v$}}$, and smoothing length $h$
(in total 7 dimensions)
and vector $\mbox{\boldmath{$X$}}_j$ ($j$-data) contains mass $m$
in addition to the same data as $\mbox{\boldmath{$X$}}_i$.
For the second stage, $\mbox{\boldmath{$X$}}_i$ contains
$\mbox{\boldmath{$r$}}$, $\mbox{\boldmath{$v$}}$, $h$, $\rho$, $c$, $P/\rho^2$
(used in equations (\ref{pf}) and (\ref{u})) 
and a quantity $f$ defined in equation (\ref{ff})
(in total 11 dimensions) and $\mbox{\boldmath{$X$}}_j$ contains mass $m$ in addition.

So far, we have made pipeline descriptions for those two stages
that is processed by PGR \citep{Hamada_2005a}.
Each of them is a simple text file of about 200 lines, and 
basically each line corresponds to a definition of one arithmetic operations.
Numbers of arithmetic operations except floating-point comparison for first stage and second stage 
are 80 and 70, respectively.
For most of operations, we have used floating point number operations.
Only exceptions are (1) the accumulation part of the pipeline where we use fixed point operations,
and (2) the kernel estimate (equation \ref{kernel}) where we also use fixed point operations.
After processing with PGR, total number of the generated VHDL source code is about 7000 lines.
Those generated VHDL source codes are synthesized using a CAD software 
provided by the vendor company of the FPGA chips into the configuration file for the processor FPGA.

\subsection{Accuracy Consideration}
In this section, we describe how to determine sufficient accuracy for our implementation.
Generally, conventional CPUs commonly used for numerical calculation
have only two selections of floating number representation 
such as single (bit-length of the fractional is 23-bit) and double precision (53-bit)
with IEEE 754 standard.
One of reasons behind great success of GRAPE project was
that they have decided not to use double precision operations
for constructing the gravity pipeline.
In implementing a logic for floating operations,
amount of needed resource (i.e., number of transistor) 
is practically proportional to the square of bid-length of the fractional.
This is absolutely true for the case of FPGA implementation and 
it is completely desirable to use as much as small bit-length of the fractional
since resource limitation of FPGA is more severe than the case of developing LSI.
A task we have to do before we really implement pipelines
is to find an optimized bit-length of the fractional
by considering how many bit-length or how much accuracy is sufficient for our application.
In PGR, we can freely select bit-length of the fractional up to 23-bit
and the exponent up to 8-bit.

\begin{figure}
\begin{center}
\scalebox{0.8}{\includegraphics{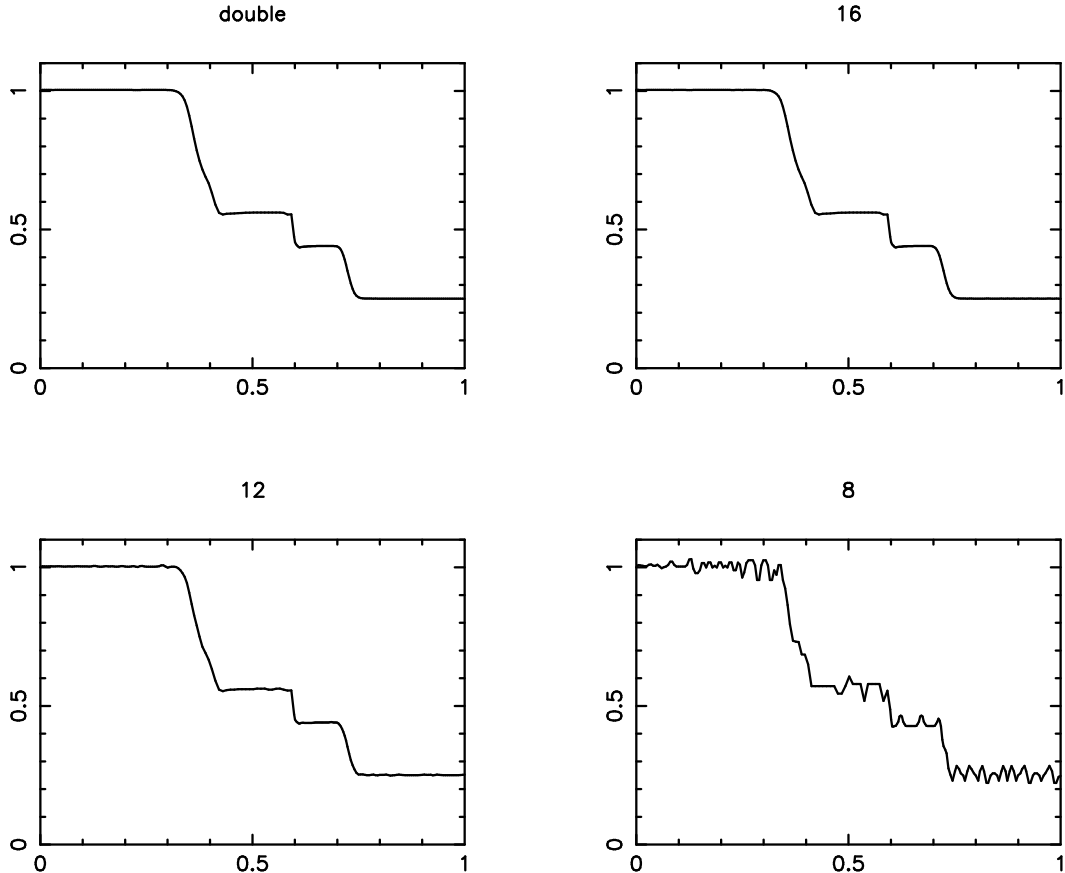}}
\caption{
Density snapshots at $t = 0.15$ for the Sod's shock tube test.
Each panel shows a result with bit-length for fraction of 
53-bit (double precision), 16-bit, 12-bit, and 8-bit, respectively 
from the upper left to the lower right.
}
\label{SHOCK}
\end{center}
\end{figure}

In the present work, to find sufficient bit-length for our application of SPH simulations, 
we have tested ability of our pipelines with various bit-length.
For this purpose, we have selected the 1-D Sod's shock tube problem \citep{Sod_1978}
as a test case.
This famous test problem is the most typical standard problem for testing 
any scheme for the Euler equation and has an analytic solution.
Ultimate goal of this test here is how the bit-length of the fractional part
affects and changes results of the Sod's shock tube problem
if compared to the results with double precision operation on the host computer.
Using PGR, we easily generate software emulators
for the SPH pipelines with desired bit-length.
We have combined those generated software emulators with our SPH code and
calculated the evolution of the Sod's shock tube problem up to $t = 0.15$.
For comparison, we have calculated the evolution with double precision operations
on the host computer using the exactly same initial condition.
In Figure \ref{SHOCK}, we show density snapshots at $t = 0.15$
with different bit-length for the fraction, 
i.e., 53-bit(double precision), 16-bit, 12-bit and 8-bit.
In all case except in the case of double precision, 
we fix the bit-length of the exponent to 8-bit and this size does not affect the results.
With 8-bit for the fraction, we apparently notice that accuracy is not enough at all
and non-physical oscillation seems to be produced due to the lack of accuracy.
However, with 12-bit or 16-bit, there is none or little such 
non-physical oscillation in the density snapshots.
Even with 12-bit, there is no distinguishable difference to the results with 
the double precision operations.

\begin{figure}
\begin{center}
\scalebox{0.8}{\includegraphics{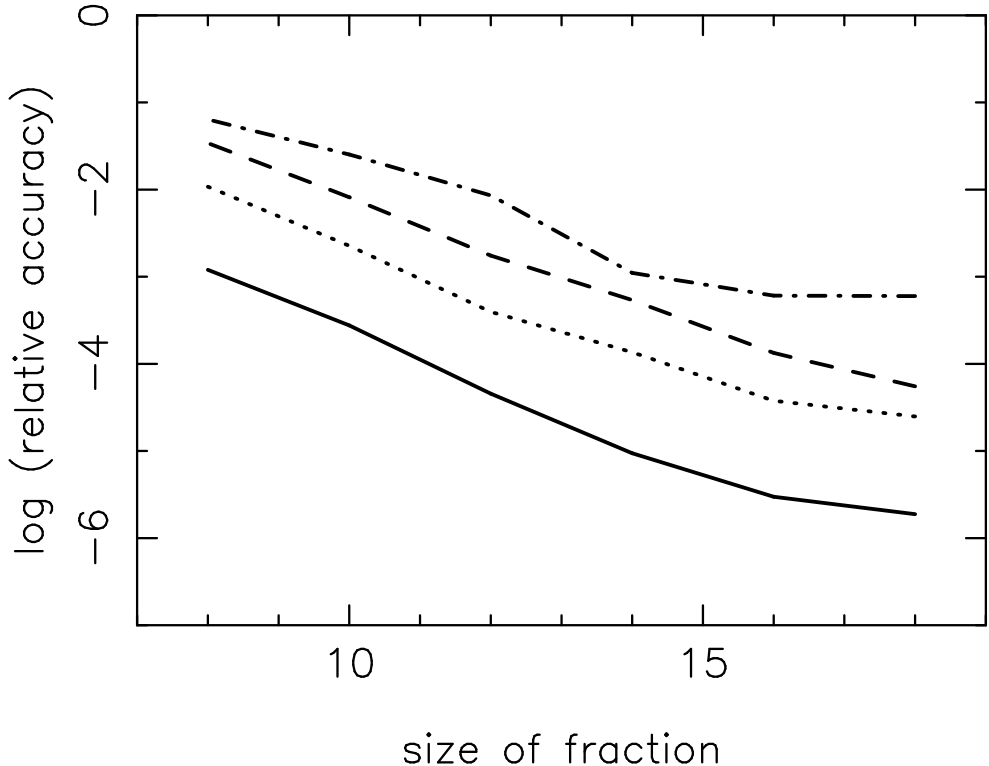}}
\caption{
The relative accuracy for position (solid line), density (dashed line),
velocity (dot-dashed line) and internal energy (dotted line) are shown as a function
of the size of the fraction.
}
\label{SHOCK2}
\end{center}
\end{figure}

To see more clearly how the size of the fraction affect the results, 
we measure and plot accuracy for different cases in Figure \ref{SHOCK2}.
Here, we define the relative accuracy for position, density, velocity and internal energy
as the average of error over all particles as
\begin{equation}
F^{i}_{\rm error} = \frac{1}{N} \sum_{i = 1}^N \frac{|F_{\rm double}^i- F_{\rm n}^i|}{|F_{\rm double}^i|},
\end{equation}
where $F_{\rm double}^i$ is the results for $i$-th particle obtained with double precision operations
and $F_{\rm n}^i$ is the results for $i$-th particle obtained with operations with $n$-bit for the fraction.
In Figure \ref{SHOCK2}, the relative accuracy for position, density, velocity and internal energy
are plotted as a function of the size of the fraction with 
the solid line, dashed line, dot-dashed line and dotted line, respectively.
More specifically, with 16-bit for the fraction, 
the relative accuracies for position, density, velocity and internal energy
are $2.98 \times 10^{-6}$, $1.33 \times 10^{-4}$, 
$6.06 \times 10^{-4}$, and $3.77 \times 10^{-5}$, respectively.
Furthermore, we compare those results with analytic solution 
of the shock tube problem and obtained following results;
(a) with double precision on HOST, 
the relative accuracies (compared to the analytic solution) for density and internal energy
are $2.556 \times 10^{-2}$ and $1.003 \times 10^{-2}$, respectively.
(b) with 16-bit for the fraction,
the relative accuracies for density and internal energy
are $2.548 \times 10^{-2}$ and $9.995 \times 10^{-3}$, respectively.
Namely, numerical error that is intrinsic to the SPH scheme and
may be intrinsic to our implementation of the SPH scheme
is much larger than numerical error caused by the reduced precision
used in the present work.

Accordingly, we adopt 16-bit for the fraction and 8-bit for the exponent in the present work.
Note this choice of 16-bit for the fraction is desirable in the sense
that the dedicated logic for signed $17 \times 17$ multiplier
in our processor FPGA chip is effectively utilized.

\subsection{Implementation}
A most distinct difference between the SPH pipelines and the gravity pipeline is 
that the SPH interaction is short-range force while the gravity is long-range force.
In the present work, we have adopted conventional spline kernel for 
our implementation of SPH pipelines, therefore, only neighbor particles in
inside of $r < 2h$ do contribute to the summations such
as equations (\ref{rho}), (\ref{v1}), (\ref{v2}), (\ref{pf}),
and (\ref{u}).
Accordingly, the first task of the conventional SPH codes running on a usual CPU 
is to make a list of neighbor particles (neighbor list),
and implementation of the neighbor search  may drastically alter
the performance of the SPH code.
Even with the SPH pipelines running on FPGA chips, 
we can have a several possibilities how to use the SPH pipelines as follows.
Here, we describe two possible algorithms developed so far for the present work.
In each algorithm, we only present simplified step-by-step algorithm.
In the followings, ``HOST'' and ``PROGRAPE'' mean the host computer and the FPGA board(s) of
the PROGRAPE-3 system, respectively.

\subsubsection*{Direct summation algorithm}
\begin{enumerate}
\item HOST prepares $j$-data for the first stage.
\item HOST sends $j$-data for the first stage to PROGRAPE.
\item For all particles, PROGRAPE calculates $\rho$, $\nabla \cdot \mbox{\boldmath{$v$}}$,
$\nabla \times \mbox{\boldmath{$v$}}$ and $n_{\rm neighbor}$, and returns the results to HOST.
\item HOST computes the equation of state and sets up $j$-data for the second stage.
\item HOST sends $j$-data for the second stage to PROGRAPE.
\item For all particles, PROGRAPE calculates $\dot {\mbox{\boldmath{$v$}}}$
and $\dot u$, and returns the results to HOST.
\item HOST integrates the Euler equation using the results obtained by the steps 3 and 6.
\end{enumerate}
As far as we set smoothing length $h$ correctly, 
calculating the summations in equations (\ref{rho}), (\ref{v1}), (\ref{v2}), (\ref{pf}), and (\ref{u})
for all particles has no problem at all.
However, this algorithm is only efficient if $N$ is less than the maximum size of $j$-data
memory in the system (in the present work, it equals to 8,192).
And $N = 8192$ is usually not a practical number of particles
if compared to current standard of SPH simulations.

\subsubsection*{Neighbor algorithm}
With the direct summation algorithm, 
we force PROGRAPE doing unnecessary operations
because typically, the average number of neighbor particles is only $\sim 30 - 100$.
To eliminate those unnecessary operations, before using PROGRAPE, 
one can construct a neighbor list on HOST and send only those particles in the neighbor
list as $j$-data.
In an algorithm combining the tree method with GRAPE
\citep{Makino_1991}, they have construct a interaction list
using the tree algorithm for a bunch of close particles
instead of each particle separately.
As a result, calculation cost on HOST for
the construction of interaction lists is reduced
by a factor of the average number of particles in those bunch ($n_g$ in their notation).
Here, we can adopt a similar approach since
those particles in a bunch, that is a group of particles very close each other, 
should have similar neighbor list.
Accordingly, it is natural to construct a shared neighbor list
for those particles in the bunch.
Here, $n_b$ is the number of bunch.
\begin{enumerate}
\item HOST constructs the tree structure for all particles.
\item HOST constructs a list of bunch.
\item For $i$ = 1 to $n_b$ (all bunch) , repeat following steps.
\begin{enumerate}
\item HOST constructs a neighbor list for $i$-th bunch.
\item HOST only sends $j$-data of the obtained neighbor particles for the first stage to PROGRAPE.
\item For particles in $i$-th bunch, PROGRAPE calculates $\rho$, $\nabla \cdot \mbox{\boldmath{$v$}}$,
$\nabla \times \mbox{\boldmath{$v$}}$ and $n_{\rm neighbor}$, and returns the results to HOST.
\end{enumerate}
\item HOST computes the equation of state and sets up $j$-data for the second stage.
\item For $i$ = 1 to $n_b$ (all bunch), repeat following steps.
\begin{enumerate}
\item HOST only sends $j$-data of the neighbor particles for the second stage to PROGRAPE.
\item For particles in $i$-th bunch, PROGRAPE calculates $\dot {\mbox{\boldmath{$v$}}}$
and $\dot u$, and returns the results to HOST.
\end{enumerate}
\item HOST integrates the Euler equation using the results obtained by the steps 3 and 6.
\end{enumerate}
In the case of the neighbor algorithm, the number of neighbor search on HOST is
$n_b \sim N/n_{\rm group}$, where $n_{\rm group}$ is the average number of particles in those bunch.
To construct a list of bunch, we use the tree algorithm as presented in \cite{Makino_1991},
and in the course of this construction, 
we recursively walk through tree nodes to see if the number of particle
in the tree code is less than $n_{\rm crit}$,
and if this condition satisfies, we make those particles in the tree node as a new bunch.
Usually, $n_{\rm group}$ is smaller than $n_{\rm crit}$ by a factor of 2.0 or so.
And, $n_{\rm crit}$ is a parameter that should be optimized
depending on the speed of HOST and PROGRAPE. 
In the present work, we have found $n_{\rm crit} = 200 - 300$ is a good choice as shown in later.

\subsection{Performance Results}
In this section, we show the performance of PROGRAPE-3 for the SPH simulations results obtained so far.
Since all operations in a pipeline generated by PGR are working in parallel
(i.e., pipelined arithmetic operation),
at every clock, our first stage pipeline for SPH calculations
executes 80 floating-point operations at the same time.
On single Bioler-3 board, we can implemented 8 pipelines, 
namely, 2 pipelines per one processor FPGA chip
with 78 \% of the resource utilization of the chip.
Although the clock frequency of the processor FPGA is arbitrary, 
our SPH pipelines are working correctly at maximum frequency of 133.3 MHz.
In this case, theoretical peak performance of the first stage pipeline
is equivalent to 
$80 {\rm (operations)} \times 2 {\rm (pipelines)} \time 4 {\rm (chips)} 133.3 \times 10^6 {\rm (Hz)} = 85.3$ GFLOPS.
In practice, there are a several reasons that limit the performance of the system such that
(1) communication time between HOST and PROGRAPE moderates the performance and 
(2) with the neighbor algorithm, time for neighbor search on HOST is another limiting factor.

\subsubsection*{Direct summation algorithm}
\begin{figure}
\begin{center}
\scalebox{0.5}{\includegraphics{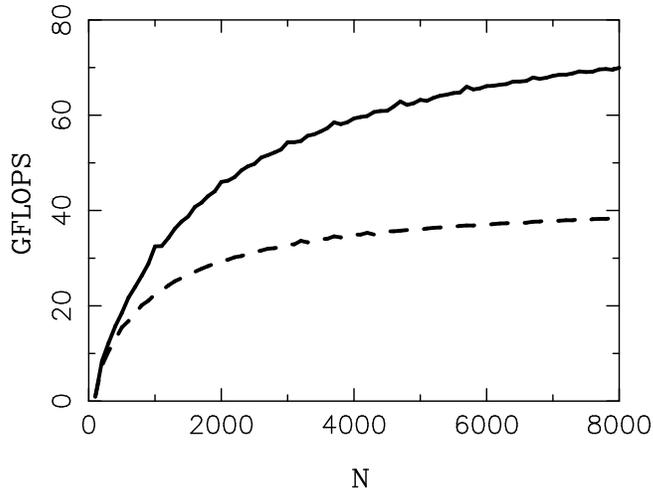}}
\caption{
Obtained performance (in GFLOPS) of the SPH first stage
as a function of $N$ are shown
in dotted line (clock speed at 66.6MHz) and solid line (133.33MHz).
}
\label{DIRECT}
\end{center}
\end{figure}

First, we present the results with ``Direct summation algorithm'' in Figure \ref{DIRECT}
In this figure, with 8 pipelines of the first stage on PROGRAPE, 
we plot the performance in GFLOPS for calculating
$\rho$, $\nabla \cdot \mbox{\boldmath{$v$}}$, $\nabla \times \mbox{\boldmath{$v$}}$ and
$n_{\rm neighbor}$
as a function of $N$ where the solid and doted lines show the results with clock frequency of 133.3 MHz
and 66.6 MHz, respectively.
As clearly shown, in either case of clock speed, 
almost 80 \% of the peak performance ($> 60$ GFLOPS for 133.3 MHz) is obtained for $N > 6000$.
Meanwhile, peak performance of double precision operation on HOST (Opteron 2.4GHz) is 4.8GFLOPS,
and obtained performance of ``Direct summation algorithm'' on HOST is $\sim 1.2$ GFLOPS.
Namely, in this case, calculations on PROGRAPE is 50 times faster than same calculations on HOST.

\begin{figure}
\begin{center}
\scalebox{0.5}{\includegraphics{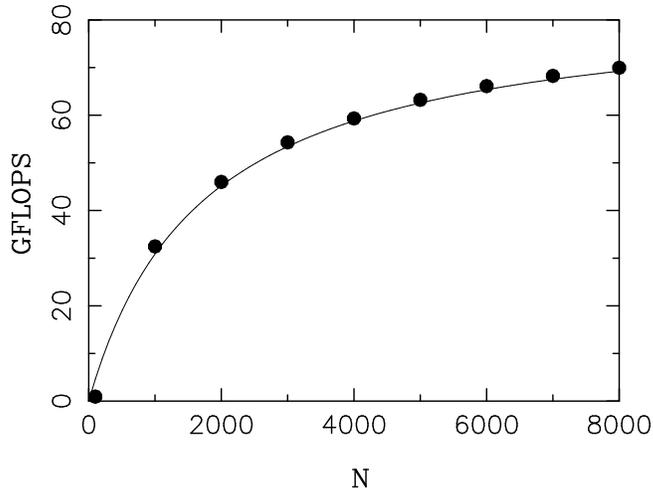}}
\caption{
Comparison of obtained performance of the SPH first stage
and our model formula.
The solid dots represent the measured results and the solid line is our model.
}
\label{DIRECT_model}
\end{center}
\end{figure}

Calculation time for ``Direct summation algorithm'' 
for the first stage on FPGA is expressed as
\begin{equation}
T_{\rm direct} = N ((n_{\rm J}+n_{\rm I}) t_{\rm write} + n_{\rm F} t_{\rm read}) + N^2 \frac{t_{\rm pipe}}{n_{\rm pipe}}
 + \frac{N}{n_{\rm pipe}} t_{\rm internal},
\end{equation}
where $n_{\rm J}$, $n_{\rm I}$, and $n_{\rm F}$ are number of byte per one particle
for $j$-data, $i$-data and $f$-data, respectively and,
$t_{\rm write}$ and $t_{\rm read}$ are transfer speed (in second per byte)
for read and write operation, respectively.
Also, $t_{\rm pipe}$ is the time spent on the pipeline for one particle,
$t_{\rm internal}$ is the time required to transfer $f$-data from the processor FPGA to the interface FPGA,
and $n_{\rm pipe}$ is the number of pipelines on the system.
In the present case of the first stage pipeline, $n_{\rm J} = n_{\rm I} = 32$,
and $n_{\rm F} = 48$, and $n_{\rm pipe} = 8$.
For $t_{\rm write}$ and $t_{\rm read}$, we have measured actual transfer time and obtained
$1/t_{\rm write} \sim 145$ MB sec$^{-1}$ and $1/t_{\rm read} \sim 42$ MB sec$^{-1}$.
Note these obtained results also include time required for conversion from/to double precision
on HOST to/from internal floating point format on PROGRAPE
and other miscellaneous task such as packing of data.
The time constant $t_{\rm pipe}$ is estimated as $t_{\rm pipe} = 1/c_{\rm pipe}$,
where $c_{\rm pipe}$ is the clock frequency of the pipeline,
and $t_{\rm internal} = n_{\rm freg} \times (2 + n_{\rm pipe})/c_{\rm bus}$,
where $c_{\rm bus} = 66.6 \times 10^6$ (Hz) is the clock frequency of the internal bus of the board.
Here, $n_{\rm freg} = 8$ is the number of registers that contain $f$-data
in each pipeline and ``2'' is a constant determined by the control logic inside the pipeline.
In Figure \ref{DIRECT_model}, we compare the measured results
for the pipeline with 133.3 MHz (also shown as the solid line in Figure \ref{DIRECT})
and our model as $(80N^2)/T_{\rm direct}$ with parameters of $c_{\rm pipe} = 133.3 \times 10^6$ (Hz).
The solid dots represent the measured results and the solid line shows our model.
Our model formula is fairly good to reproduce the measured performance.

\subsubsection*{Neighbor algorithm}
\begin{figure}
\begin{center}
\scalebox{0.5}{\includegraphics{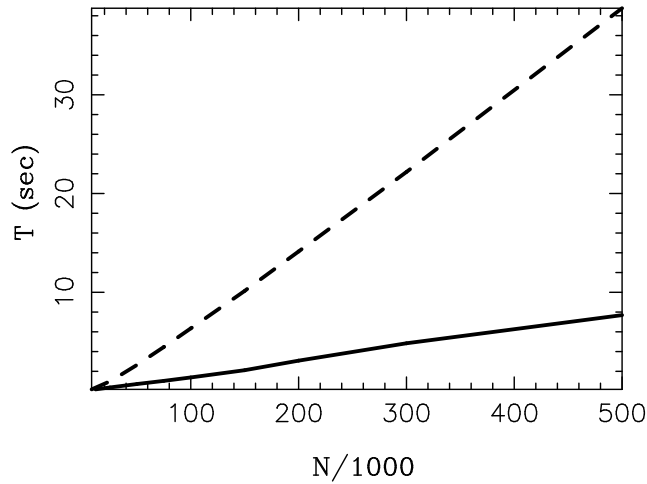}}
\caption{
The measured time for one step with and without PROGRAPE as a function of $N$.
The solid line represents the results using PROGRAPE using ``Neighbor Algorithm ''
and the dashed line shows the results without PROGRAPE.
}
\label{NB}
\end{center}
\end{figure}

In practice, ``Direct summation algorithm'' is useful only when
the number of particles is small because with this algorithm,
large fraction of operations on PROGRAPE is unnecessary
when the number of particles is large.
Here, as practical measure, we show the performance results with ``Neighbor algorithm''.
The particle distribution used for this performance test is
the initial particle distribution of so-called ``Cold Collapse Test''
\citep{Evrard_1988, Hernquist_1989}, and has the density profile of $\rho \propto r^{-1}$.
Here, we measure the required time for one step using PROGRAPE with ``Neighbor algorithm''
and show the results as a function of $N$ in Figure \ref{NB} with the solid line.
As a comparison, the dashed line presents the required time for one step
using our SPH code on the same HOST.
Here, {\it one step} is the time spend on full force calculation for all particles.
In both cases, we set the number of neighbor particle is $\sim 40$.
Moreover, in the test for PROGRAPE, we set $n_{\rm crit} = 100$ and 
obtain $n_{\rm group} \sim 30$ and the average number of $j$-particle ($\bar{n}_j$) $\sim 300$
where those numbers are slightly depending on $N$.
It is clear from this figure that the results with PROGRAPE outperform 
the results with HOST by a factor of 5, for example, 
when $N = 500,000$, PROGRAPE takes $7.61$ seconds for one step while HOST takes 38.8 seconds.

\begin{figure}
\begin{center}
\scalebox{0.5}{\includegraphics{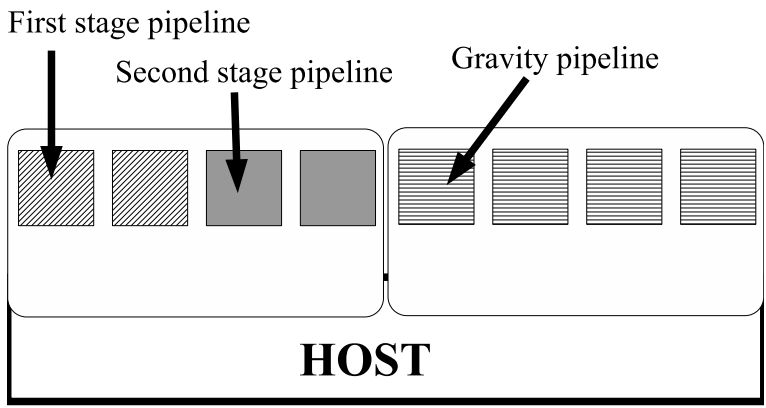}}
\caption{
The configuration used for ``Neighbor algorithm''.
In this case, two processor FPGA chips are used for each first and second stage,
respectively and thus the number of pipelines for each stage is 4.
A spare board can be used for another purpose such as for gravity interaction.
}
\label{config1}
\end{center}
\end{figure}

\begin{figure}
\begin{center}
\scalebox{0.5}{\includegraphics{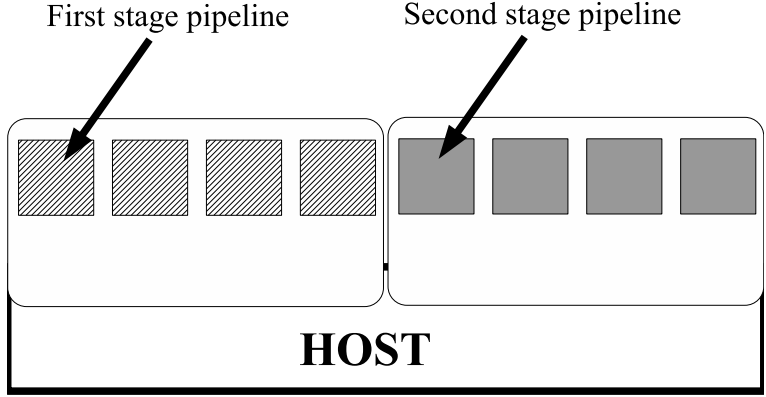}}
\caption{
The configuration used for ``Direct summation algorithm''.
In this case, all four processor FPGA chips are used for the first stage
and the number of pipelines for the first stage is 8.
Also a spare board can be used for the second stage and
this configuration is suitable for pure hydrodynamical simulations.
}
\label{config2}
\end{center}
\end{figure}

Note for those tests with PROGRAPE shown in this section,
we use two FPGA chips on one PROGRAPE
for the first stage and other two chips for the second stage
as shown in Figure \ref{config1}.
In other words, $n_{\rm pipe}$ for both stages is 4
and consequently the theoretical peak performance for both stages 
is $\sim 40$ GFLOPS.
On the other hand, the performance results for ``Direct summation algorithm''
are obtained with $n_{\rm pipe} = 8$, i.e., we use four FPGA chips on PROGRAPE
for the first stage as shown in Figure \ref{config2}.
Although we are free to use any number of boards for our tests or applications, 
the configuration in Figure \ref{config1} 
such that we use two PROGRAPE boards and
one PROGRAPE is used for SPH interaction and other for gravity interaction
is realistic at the moment.
This is because our main application of the technique presented in the present paper
will be simulations of the galaxy evolution that involves both hydrodynamics and gravity.
If one wants to calculate pure hydrodynamical simulations, 
the configuration shown in Figure \ref{config2} may be better choice.
Even in this case, one can add a third board (PROGRAPE or any GRAPE) 
to this configuration for gravity interaction.

To see how the required time for one step depends on $n_{\rm crit}$,
we have done the same test above with different $n_{\rm crit}$.
Table \ref{NB_comp} shows the comparison between those results for $N = 500,000$.
In this case, with $n_{\rm crit} = 500$, we obtain best performance and the actual
average number of particles in the bunch ($n_{\rm group}$) is $\sim 175$
with $\bar{n}_j \sim 681$.
Although this small number of $\bar{n}_j$ is somewhat surprising for us,
this fact directs us toward next generation of PROGRAPE system with more improved 
transfer speed between HOST and PROGRAPE as explained with a following model.

The required time for one step with ``Neighbor algorithm'' is expressed as
\begin{equation}
T = T_{\rm host} + T_{\rm PROGRAPE} + T_{\rm comm},
\end{equation}
where $T_{\rm host}$, $T_{\rm PROGRAPE}$ and $T_{\rm comm}$ are the time spent on HOST,
the time spent on PROGRAPE, and the time required to transfer data between HOST and PROGRAPE, respectively.
The time $T_{\rm host}$ is expressed as
\begin{equation}
T_{\rm host} = t_{\rm tree} + \frac{N}{n_{\rm group}} t_{\rm neighbor} + t_{\rm misc},
\end{equation}
where $t_{\rm tree}$ is the time required to construct the tree structure
and make a list of bunch, and $t_{\rm neighbor}$ is 
the time required to construct a neighbor list for a bunch.
Also, $t_{\rm tree}$ depends on $N {\rm log} N$ and
$t_{\rm neighbor}$ weakly depends on $N$
and $t_{\rm misc}$ is the time required for other miscellaneous calculations on HOST.
The time $T_{\rm PROGRAPE}$ is expressed as
\begin{equation}
T_{\rm PROGRAPE} = \frac{N}{n_{\rm pipe}} \bar{n}_j (t_{\rm pipe 1st} + t_{\rm pipe 2nd})
 + 2 \frac{N}{n_{\rm pipe}} t_{\rm internal},
\end{equation}
where $t_{\rm pipe 1st}$ and $t_{\rm pipe 2nd}$ are 
the time spent on the pipeline for one particle in the first and second pipelines, respectively.
And, $T_{\rm comm}$ is expressed as
\begin{equation}
T_{\rm comm} = \left[ \frac{N}{n_{\rm group}} \bar{n}_j (n_{\rm J 1st}
+ n_{\rm J 2nd}) + N (n_{\rm I 1st} + n_{\rm I 2nd}) \right] t_{\rm write} +
		N (n_{\rm F 1st} + n_{\rm F 2nd}) t_{\rm read},
\end{equation}
where $n_{\rm J 1st}$ and $n_{\rm J 2nd}$ are the number of byte per one particle 
for $j$-data of the first and second stages, respectively, and 
$n_{\rm I 1st}$, $n_{\rm I 2nd}$, $n_{\rm F 1st}$, and $n_{\rm F 2nd}$ are same notation.

For simplicity, we approximate the time $T_{\rm host}$ as a linear function of $N$.
From the measurement, we use the relation $T_{\rm host} = \alpha_{\rm host} N$ in the following
where $\alpha_{\rm host} \sim 2.7 \times 10^{-6}$ (sec) for the present HOST.

\begin{figure}
\begin{center}
\scalebox{0.5}{\includegraphics{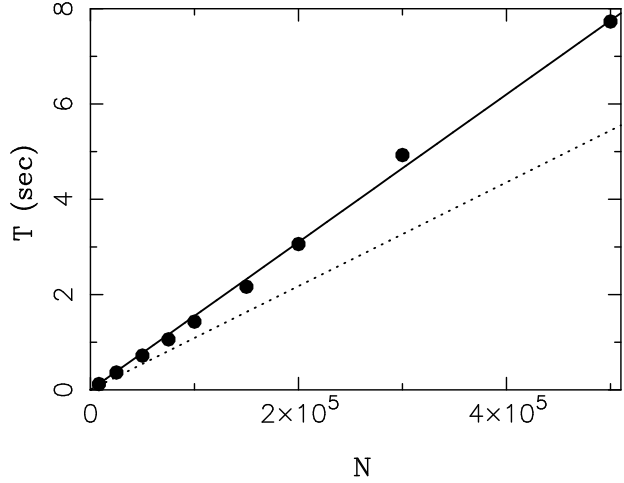}}
\caption{
The solid dots represent the measured time ($T = T_{\rm host} + T_{\rm PROGRAPE} + T_{\rm comm}$)
for one step with ``Neighbor Algorithm (B)'' as a function of $N$.
The solid line shows our model formula explained in the text.
To see the fraction of the time spent on data transfer between HOST and PROGRAPE,
the dotted line represents only $T_{\rm comm}$.
}
\label{NB_model}
\end{center}
\end{figure}

In Figure \ref{NB_model}, we compare the measured results for
$T = T_{\rm host} + T_{\rm PROGRAPE} + T_{\rm comm}$ and our model presented above. 
The solid dots represent the measured results and the solid line shows our model.
The parameters for our model in the present case is as follows;
$t_{\rm pipe 1st} = t_{\rm pipe 2nd} = 1/c_{\rm pipe}$ where $c_{\rm pipe} = 133.3 \times 10^6$ (Hz).
The time $t_{\rm internal}$ is the same as explained in section ``Direct summation algorithm''.
We set $n_{\rm pipe} = 4$ since we use two FPGA chips for each stage (see Figure \ref{config1}).
We set $n_{\rm J 1st} = 32, n_{\rm I 1st} = 32, n_{\rm F 1st} = 64$, 
$n_{\rm J 2nd} = 40, n_{\rm I 2nd} = 44$, and $n_{\rm F 2nd} = 64$.
For $t_{\rm write}$ and $t_{\rm read}$, we have measured actual transfer time
and obtained the effect transfer rates between HOST and PROGRAPE as 
$1/t_{\rm write} \sim 145$ MB sec$^{-1}$ and $1/t_{\rm read} \sim 37$ MB sec$^{-1}$.
Finally, we set $\bar{n}_j = 350$ and $n_{\rm group} = 25$ for the present case.
For most of the data points, our model formula matches very well with the measured time.

In the present case of ``Neighbor algorithm'', 
parameters that have room for improvement are $t_{\rm write}$ and $t_{\rm read}$.
This is clearly shown in the doted line in Figure \ref{NB_model},
that represents the time required for data transfer between HOST and PROGRAPE.
Fairly large fraction ($\sim 70$ \%) of the time is spent on the data transfer.
This means that $n_{\rm pipe}$ (i.e., number of pipelines)
and $t_{\rm pipe}$ (i.e., clock speed of the pipeline) are not important parameters
to optimize or enhance for ``Neighbor algorithm''.
Especially, $t_{\rm read}$ is rather large in the present configuration, namely,  
effective transfer speed from PROGRAPE to HOST ($\sim 40$ MB sec$^{-1}$)
is much slower than expected from the theoretical transfer rate of the board ($512$ MB sec$^{-1}$)
because the design of the present transfer logic is not optimal.

\begin{figure}
\begin{center}
\scalebox{0.5}{\includegraphics{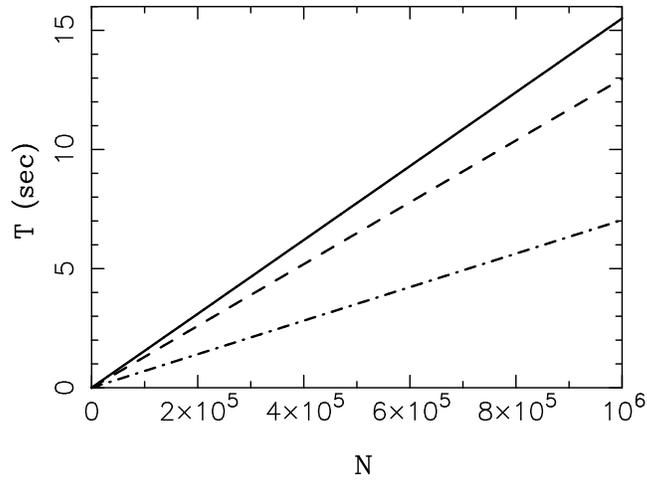}}
\caption{
We compare predicted $T = T_{\rm host} + T_{\rm PROGRAPE} + T_{\rm comm}$ with ``Neighbor Algorithm (B)''
as a function of $N$ for several different data transfer speed between HOST and PROGRAPE.
The solid line corresponds to the present configuration of PROGRAPE-3 system ({\b Model P3}).
The dashed line and dot-dashed line show 
the prediction with only enhanced read data transfer speed
and with enhanced both read and write data transfer speed, respectively.
}
\label{NEW_board}
\end{center}
\end{figure}

To see how $t_{\rm write}$ and $t_{\rm read}$ affect the performance, 
we plot our model formulae for a several different cases in Figure \ref{NEW_board}.
The solid line represents the required time for the present configuration (Model P3).
If we assume $t_{\rm read}$ equals to $t_{\rm write}$ of the configuration,
namely we set $1/t_{\rm write} = 1/t_{\rm read} = 145$ MB sec$^{-1}$, 
the result is shown as the dashed line.
In this case, the performance is $\sim 20$ \% better than Model P3.
Furthermore, the dashed line shows the results with
$1/t_{\rm write} = 1/t_{\rm read} = 500$ MB sec$^{-1}$, that is 
much higher data transfer rate than the present Bioler-3 board.
In this case, the performance is $\sim 200$ \% better than Model P3
and the one step for 1 million particles SPH simulation takes $7$ seconds with the present HOST.
Those assumed data transfer rate is not out of reach
if we consider new I/O standard such as PCI-Express.
With PCI-Express, the theoretical data transfer rate for one-lane configuration
is $250$ MB sec$^{-1}$, and four/eight-lane configuration has 
the theoretical rate of $1000/2000$ MB sec$^{-1}$.
For future development of new boards and optimization of those boards, 
it is very crucial to consider the data transfer between HOST and PROGRAPE.

\begin{table}
\begin{center}
\begin{tabular}{c|c|c|c}
$n_{\rm crit}$ & $n_{\rm group}$ & $\bar{n}_j$ & time\\
50 & 19.6 & 256 & 8.3 \\
100 & 31.1 & 320 & 7.6 \\
200 & 42.3 & 363 & 7.3 \\
500 & 175  & 681 & 6.6 \\
1000 & 263 & 829 & 6.8 \\
\end{tabular}
\end{center}
\caption{
Comparison between results with different $n_{\rm crit}$.
}
\label{NB_comp}
\end{table}

\subsection{Application Tests}
\begin{figure}
\begin{center}
\scalebox{0.75}{\includegraphics{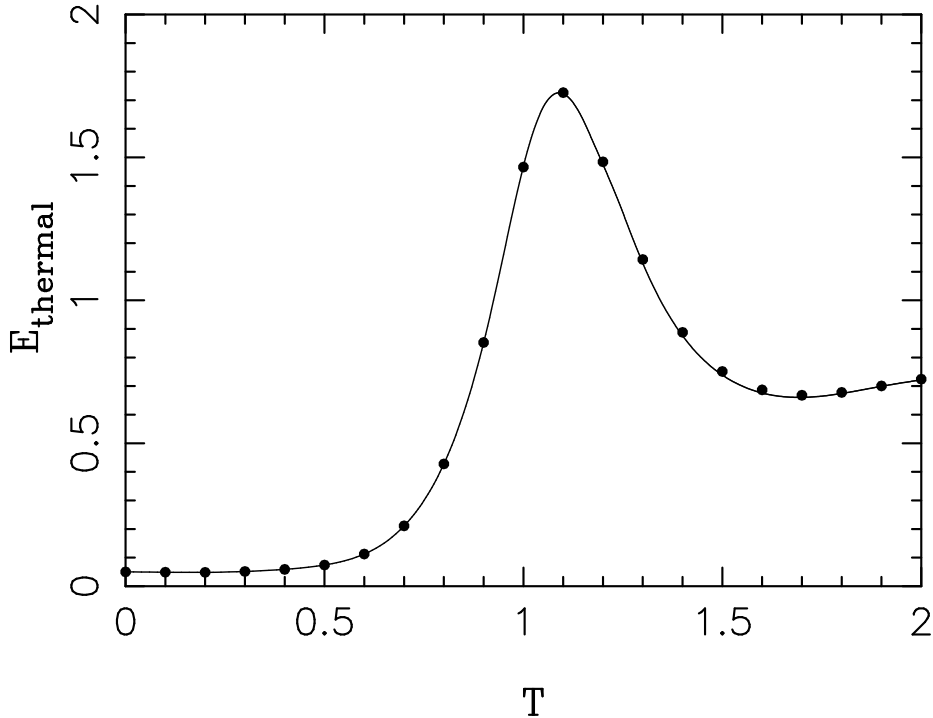}}
\caption{
The evolution of thermal energy for the Cold Collapse Test with $N = 25,000$.
The solid dots show the results with double precision operations on HOST and 
the thin solid line shows the results with PROGRAPE.
}
\label{energy}
\end{center}
\end{figure}

In this section, we present the results for a test application using PROGRAPE.
The test application is a standard test of SPH codes, 
so-called ``Cold Collapse Test'' \citep{Evrard_1988, Hernquist_1989}.
This test has been used by many authors to see basic abilities of their 
SPH simulation code for astrophysical problem involving self gravity.
In this problem, we follow the evolution of collapse of a gas sphere
with the density profile of $\rho \propto r^{-1}$.
The problem setup is similar to structure formation in
cosmological simulations.
The actual density profile of the problem is as follows;
\begin{equation}
\rho(r) = \frac{M}{2 \pi R^2} \frac{1}{r},
\end{equation}
where $M$ is the total mass of the sphere and $R$ is the radius of the sphere.
Total internal energy of the sphere is $E_{\rm thermal} = 0.05 M/R$.
We set $M = R = 1$ in the present work and initially,
total potential energy of the sphere ($|E_{\rm pot}| \sim 0.6$) is much larger than 
total internal energy of the sphere ($E_{\rm thermal} = 0.05$).
Accordingly, the sphere quickly collapse and shock is produced due to the collapse.
In Figure \ref{energy}, we show the evolution of $E_{\rm thermal}$ for two cases;
(1) calculation with double precision operations on HOST 
and (2) calculation using PROGRAPE with Neighbor algorithm (B).
In both case, initial particle distribution is same and $N = 25000$
and we use the tree method on HOST for gravity calculation with 
Barnes-Hut criterion $\theta = 0.75$.
In this Figure, the solid line shows the results corresponding to case (2)
(with a configuration shown in Figure \ref{config1} but do not use gravity pipelines for this case)
and the solid dots present the results with case (1).
There is no significant deviation between two results.
This means that SPH calculation on PROGRAPE with the fractional 16 bit is
accurate enough to follow shock wave produced by collapse of cold gas sphere.

The performance of this test is much encouraging for us as shown in table \ref{CC_time}.
In this table, we compare the averaged time (in sec) for one step with 4 different cases;
(1) all calculation with double precision operations on HOST (shown in 2nd column),
(2) SPH calculation on PROGRAPE and gravity calculation with the tree method on HOST (3rd column),
(3) SPH calculation on PROGRAPE and gravity calculation on PROGRAPE with the direct summation scheme (4th column),
and (4) SPH calculation on PROGRAPE and gravity calculation on PROGRAPE with the tree method (5th column).
The 1st column shows the number of particles used in those performance tests.
The numbers inside parenthesis in the 3rd to 5th column show speedup factor compared to
the results in 2nd column.
We use exactly the same initial condition for ``Cold Collapse Test'' that is already described.
In the case (3) and (4), we use a configuration shown in Figure \ref{config1}.
For a reference, the peak performance of gravity pipelines running on PROGRAPE for those tests
is 324 GFLOPS \citep[see some preliminary performance results of our gravity pipelines][]{Hamada_2005a}.
The performance results with PROGRAPE is at least 4 times faster and in some case 11 times faster than
the calculation on HOST.
Namely, combination of SPH pipelines running on PROGRAPE and GRAPE or PROGRAPE with gravity pipelines
offers us very promising performance measures.
This clearly shows for the first time that using considerable computing power offered
by a hardware we can accelerate the speed of SPH simulations of a simple astrophysical phenomena.
Our results open new and extensive possibility of hardware acceleration of complicated
and computing intensive applications using PROGRAPE architecture or similar approach.

\begin{table}
\begin{center}
\begin{tabular}{c|c|c|c|c}
SPH & HOST & PROGRAPE & PROGRAPE & PROGRAPE \\
Gravity & HOST & HOST(tree) & PROGRAPE(direct) & PROGRAPE(tree) \\
\hline
$N=25,000$  & 2.30 & 1.39 ({\bf 1.6}) & 0.54 ({\bf 4.3}) & 0.49 ({\bf 4.7}) \\
$N=50,000$  & 6.63 & 3.35 ({\bf 2.0}) & 1.30 ({\bf 5.1}) & 0.97 ({\bf 6.8}) \\
$N=100,000$ & 17.8 & 7.96 ({\bf 2.2}) & 3.47 ({\bf 5.1}) & 1.95 ({\bf 9.1}) \\
$N=500,000$ & 107  & 45.7 ({\bf 2.3}) & 51.9 ({\bf 2.1}) & 9.62 ({\bf 11.1}) \\
\end{tabular}
\end{center}
\caption{
In this table, we show the required time (in sec) for one step with 4 different cases;
(1) all calculation with double precision operations on HOST (shown in 2nd column),
(2) SPH calculation on PROGRAPE and gravity calculation with the tree method on HOST (3rd column),
(3) SPH calculation on PROGRAPE and gravity calculation on PROGRAPE with the direct summation scheme (4th column),
and (4) SPH calculation on PROGRAPE and gravity calculation on PROGRAPE with the tree method (5th column).
The numbers inside parenthesis in the 3rd to 5th column show speedup factor compared to
the results in 2nd column.
}
\label{CC_time}
\end{table}

\section{Summary}
After the introduction of the concept of the GRAPE architecture, 
subsequent development of the GRAPE systems enable us to quite further 
advance the number of particles or size of gravitational many-body simulations.
Especially, in collisional evolution of dense stellar cluster, 
those acceleration of simulation obtained by GRAPE-4 and GRAPE-6 system
has played a vital role to understand a several important physical consequence.
As a natural advance of the GRAPE architecture, 
\cite{Hamada_2000} has proposed the PROGRAPE architecture and 
shown the first development system of PROGRAPE-1 board.
In the PROGRAPE system, we have used FPGA chips as a computing engine
instead of specially developed LSI chips as used in all GRAPE project.
Programmability and flexibility of FPGA make us possible to implement
arbitrary particles interaction expressed as equation (\ref{basic}).

In this paper, after briefly introduce PROGRAPE-3 system, 
which is our third generation of PROGRAPE architecture, 
we present a novel approach to accelerate astrophysical hydrodynamical simulations.
PROGRAPE-3 consists of the Bioler-3 board that has four large FPGA chips and a host computer.
On PROGRAPE-3 system, we implement logic circuits (numerical pipeline) for
gravitational pipeline that is identical to GRAPE-5 system
and force pipeline for the SPH method.
Generally, implantation of such pipelines for astrophysical simulations
needs complex development efforts and considerable development time.
To reduce those complex efforts, 
we have developed a software system (PGR system) that enable us to easily implement
arbitrary particle interaction such as equation (\ref{basic}).

With PGR, we implement SPH calculation on PROGRAPE-3 system
and have shown for the first time
that the calculation speed of SPH simulations can be accelerated with our novel approach. 
Due to its complexity, a development of special LSI for SPH calculations 
is much harder than the development of the GRAPE chips.
However, the combination of PGR and a computational board based on  
PROGRAPE architecture easily enable us to develop numerical pipelines
for SPH calculations on FPGA.
For implementing SPH calculations on FPGA, 
we did a simple test to see how numerical accuracy affects 
results of SPH calculation by changing bit-width of the fractional of floating point operations.
We found that 16-bit for the fraction is enough accurate for instance.
With this obtained numerical accuracy, the peak performance of SPH pipelines 
running on PROGRAPE-3 system is $\sim 85$ GFLOPS. 
This performance number is almost 20 times faster than the peak performance of HOST.
Finally, we did a realistic performance test and obtained that
the SPH calculation using PROGRAPE-3 board is 5 times faster than 
the calculation on HOST.
Currently, the performance of the SPH calculation on the PROGRAPE system
is mainly limited by the data transfer speed between HOST and PROGRAPE
and can be much better if we could have a new board with much faster interface.
Our results clearly show that numerical calculations
on FPGA is a very promising way to enhance the performance of SPH simulations.
The methodology that we have presented in this paper will be also directly applicable 
to other compute intensive calculations on numerical astrophysics and computational physics.

The authors would like to thank T. Ebisuzaki and J. Makino for useful discussions.
The authors (N.N. and T.H.) specially thanks
R. Spurzem, R. M\"{a}nner and other colleagues
at the Astronomisches Rechen-Institut and University Mannheim
for the hospitality during our stay in both institutes
where a part of this paper has been written.
A part of the work has been supported by the Exploratory Software Project
2004, 2005 of Information Technology Promotion Agency, Japan.

\bibliography{text}

\end{document}